\documentclass[aip,jcp,amsmath,amssymb,reprint]{revtex4-1}

\usepackage{graphicx,graphics,times,bm,bbm,bbold,amssymb,amsfonts,dsfont,xcolor,color,cancel,hyperref,tabularx}
\usepackage[normalem]{ulem}
\usepackage{braket}
\usepackage{comment}
\hypersetup{colorlinks,linkcolor={blue},citecolor={blue},urlcolor= {blue}}
\let\oldsqrt\sqrt
\def\sqrt{\mathpalette\DHLhksqrt}
\def\DHLhksqrt#1#2{%
\setbox0=\hbox{$#1\oldsqrt{#2\,}$}\dimen0=\ht0
\advance\dimen0-0.2\ht0
\setbox2=\hbox{\vrule height\ht0 depth -\dimen0}%
{\box0\lower0.4pt\box2}}
\DeclareFontFamily{OT1}{pzc}{}
\DeclareFontShape{OT1}{pzc}{m}{it}%
              {<-> s * [1.25] pzcmi7t}{}
\DeclareMathAlphabet{\mathpzc}{OT1}{pzc}%
                                 {m}{it}
   \newcommand{\Balpha}{\mbox{$\hspace{0.12em}\shortmid\hspace{-0.62em}\alpha$}}
    
\usepackage{amsmath}
\usepackage{etoolbox}
\makeatletter
\def\align@preamble{%
   &\hfil
    \strut@
    \setboxz@h{\@lign$\m@th\displaystyle{##}$}%
    \ifmeasuring@\savefieldlength@\fi
    \llap{\set@field}%
    \tabskip\z@skip
   &\setboxz@h{\@lign$\m@th\displaystyle{{}##}$}%
    \ifmeasuring@\savefieldlength@\fi
    \rlap{\set@field}
    \hfil
    \tabskip\alignsep@
}

\begin{document}
\title{Nonequilibrium mixture dynamics: A model for mobilities and its consequences}
\author{Maryam Akaberian}
\affiliation{Institut für Theoretische Physik, Georg-August-Universität Göttingen, 37077 Göttingen, Germany}
\author{Filipe C Thewes}
\affiliation{Institut für Theoretische Physik, Georg-August-Universität Göttingen, 37077 Göttingen, Germany}
\author{Peter Sollich}
\affiliation{Institut für Theoretische Physik, Georg-August-Universität Göttingen, 37077 Göttingen, Germany}
\affiliation{King's College London, Department of Mathematics, Strand, London WC2R 2LS, U.K.\\}
\author{Matthias Krüger}
\affiliation{Institut für Theoretische Physik, Georg-August-Universität Göttingen, 37077 Göttingen, Germany}
\date{\today}
\begin{abstract}
Extending the famous Model B for the time evolution of a liquid mixture, we derive an approximate expression for the mobility matrix that couples the different mixture components. This approach is based on a single component fluid with particles that are artificially grouped into separate species labelled by ``colors''. The resulting mobility matrix depends on a single dimensionless parameter, which can be determined efficiently from experimental data or numerical simulations, and includes existing standard forms as special cases. 
We identify two distinct mobility regimes, corresponding to collective motion and interdiffusion, respectively, and show how they emerge from the microscopic properties of the fluid. As a test scenario, we study the dynamics after a thermal quench, providing a number of general relations and analytical insights from a Gaussian theory. Specifically, for  systems with two or three components,  analytical results for the time evolution of the equal time correlation function compare well to results of Monte Carlo simulations of a lattice gas.
A rich behavior is observed, including the possibility of transient fractionation.
\end{abstract}

\maketitle
\section{ Introduction}

A general description of the static properties of fluids is one of the successes of statistical physics in general and density functional theory in particular~\cite{evans79,hansen2013,eschrig2003,cohen2012}. In contrast, the prediction of dynamical timescales as a direct result of the underlying kinetics remains a challenge, in spite of significant advances in the past decades~\cite{Dean1996,Kim2014,Spohn1991,BettoloMarconi2000,SchmidtBraderJCP_2013_power_func,Scacchi_2016}. These timescales depend not just on thermodynamic driving but also on transport coefficients, which can be highly non-trivial, for example at high densities where crowding effects may slow down the system resulting in glassy behavior~\cite{gotze2009,Janssen2018}.

Transport coefficients appear generally in linear response relations~\cite{kubo1957} between thermodynamic driving forces and the response of the system~\cite{hansen2013,kardar2007,Kruger2017}. The Green-Kubo relation~\cite{green1954,kubo1957} between such transport coefficients and equilibrium time-dependent correlation functions is a  celebrated result in this context. The time-dependent correlations can be obtained, e.g.\ via the many-body Smoluchowski equation~\cite{dhont1996,Fuchs2002}, though this usually requires approximate treatments like expansions around the low density limit \cite{dhont1996}. Conversely, at high densities, when crowding effects are relevant, Mode Coupling Theory~\cite{gotze2009,Janssen2018} yields good approximations. 

The time evolution of the density field for fluids with a conserved number of particles is described by the famous model B~\cite{hohenberg1977,Bray1994,bray2002theory}, which  combines linear response theory with mass conservation and involves the mass transport coefficient as the key kinetic quantity. Due to the conserved field such fluids exhibit long-ranged fluctuations in non-equilibrium scenarios \cite{grinstein1990,spohn1983,dorfman1994,evans1998_2} and show non-equilibrium fluctuation-induced forces when quenched \cite{Rohwer2017,Rohwer2018} or exposed to temperature gradients \cite{kirkpatrick2015,kirkpatrick2016}. Here, a firm understanding of the transport coefficients becomes of great importance, as possible crowding effects can interfere and dictate the timescales on which these (non-equilibrium) fluctuations propagate in the system.

Given its importance for the kinetic description of fluids the mass transport coefficient, also so-called mobility $L$, has been investigated from several perspectives. The Dean-Kawasaki equation~\cite{Dean1996,kawasaki1998}, for example, predicts a linear dependence of the mobility on density, but this is done within a formal time evolution equation for the exact particle number density, which consists of a sum of delta-functions.

Earlier, extensive work by Batchelor on hard spheres and hard sphere mixtures~\cite{Batchelor1976,Batchelor1983} had provided expressions for the mobility as a function of particle sizes as well as density. However, these are obtained by an expansion in particle density and can hardly be used to study crowding effects. To overcome this problem, a number of models have been developed where crowding effects are added phenomenologically by including a critical density for which the mobility  vanishes~\cite{Kob1993,Fernandes2003,Levin2001}. A comprehensive microscopic understanding of crowding effects on the mobility is therefore still lacking. 

The challenges around our understanding of the mobility multiply when moving from single component fluids to multi-component systems due to the larger number of conserved fields, one for each component density. Already the equilibrium phase behaviour is more complex here due to the possibility of fractionation~\cite{Evans1998,Sollich2002,Zwicker2022},
where different mixture components interdiffuse and thus demix.
For a given initial distribution of components, fractionation leads to the formation of daughter phases with in general different composition, whose coexistence at equilibrium at fixed total density and temperature can be established using the double tangent construction~\cite{Sollich2002}. The kinetics also become substantially more complex.
According to the so-called Warren scenario~\cite{Warren1998}, one has to distinguish here relaxation of the local composition or equivalently of the relevant moment-densities~\cite{Sollich2002,Warren1998,Warren1999} on the one hand, and relaxation of the total density on the other. Following Warren, the latter can be achieved relatively easily by collective motion of particles, while the former occurs on typically longer timescales by 
interdiffusion of particles from different mixture components. Crowding primarily affects interdiffusion so this separation of timescales can become pronounced at high densities, leading for example to two-stage relaxation processes~\cite{Warren1999,castro2018,castro2019}.
More broadly, both theory~\cite{Warren1999,Grodon2007,Jacobs2013} and simulations~\cite{Wilding2006,Jacobs2013,Jacobs2021,Zwicker2022Evolved} show a fascinating range of behavior in multi-component mixtures. This is of particular interest also in biophysics~\cite{berry2018,yu2004}, where phase separation plays a crucial role in the formation of intracellular structures~\cite{alberti2017,hyman2014}.

For multi-component systems, the model B description of the time evolution of the density fields becomes a set of partial differential equations that are coupled both by thermodynamic effects and by the mobility, which is now a matrix $\mathbb{L}$. Often this matrix is approximated as diagonal~\cite{Shrinivas2021}, i.e.\ $L_{ij}\propto \rho_i\delta_{ij}$, or to follow the form for an ideal mixture with only volume exclusion~\cite{Mao2019} $L_{ij}\propto \rho_i(1-\rho_j)$, as obtained e.g.\ in polymer mixtures~\cite{Pagonabarraga2003} or in the multi-component symmetric exclusion process~\cite{Vanicat2017}. These expressions are valid either in specific regimes in parameter space, e.g.\ low densities, or rely strongly on the underlying model. Similarly to the single fluid case, a general description of the mobility matrix in multi-component mixtures is still lacking in the literature.

Keeping in mind the richness of behavior and the interplay between thermodynamic and kinetic effects in fluids with many components, we aim to obtain in this work a coarse-grained description of the mobility matrix $\mathbb{L}$ resulting from microscopic properties of the system. This is based on a model of a single component fluid that we transform into a mixture by painting particles with different ``colors'' without changing their physical properties. This model yields a general expression for $\mathbb{L}$ that exhibits two competing modes of motion: collective and interdiffusion. We show how previous models for mobilities are recovered in specific limits, which are reached by imposing certain kinetic constraints on the system, and how they favor one mode of motion or the other.

In order to test our results in a non-equilibrium fluid mixture scenario, we consider (mild) quenches from a higher to a lower temperature. We obtain a closed form expression for the time evolution of the correlation matrix, which as a dynamical quantity involves the mobility matrix. We then compare our theoretical predictions with numerical simulations of a multi-component lattice gas and confirm our main hypothesis by showing that our form of the mobility matrix yields very good approximations to the simulated time evolution of the correlation matrix for nontrivial mixtures.

The present work is structured as follows. The main equations of motion and thermodynamic quantities are introduced in Sec.~\ref{sec:kinetics}.  We introduce the painted particle model and explore its consequences in Sec.~\ref{sec:painted}. The result is an explicit expression for the mobility matrix. Sec.~\ref{sec:quench} investigates the dynamics after a quench using the mobility obtained in the previous section, and discusses a simple paradigmatic case that emphasises the mobility effects in this scenario. Finally, in Sec.~\ref{sec:simul} we describe our lattice gas simulations and compare the simulation results for correlations after a quench with our theoretical predictions. We summarize and give a brief outlook in Sec.~\ref{sec:concl}.

\section{Kinetics of multicomponent mixtures}
\label{sec:kinetics}
We start by defining a mixture of total particle number  $N$ in a volume $V$ with $N_i$ particles of species $i=1\dots M$. The density fluctuation field of species $i$, at position $\boldsymbol x$ and time $t$, is given by \cite{hansen2013}
\begin{equation}
    \phi_i(\boldsymbol x, t) = \sum_{k=1}^{N_i} \delta(\boldsymbol x - \boldsymbol x_k(t)) - \frac{N_i}{V}.
\end{equation}
where $\boldsymbol x_k$ is the position of particle $k$, one of the particles of species $i$. Due to conservation of particle number of each species in the system (we exclude chemical reactions), the dynamics of a multi-component mixture is described by a continuity equation of the form

\begin{equation}
\begin{split}
    \dot\phi_i(\boldsymbol{x},t) = \nabla\cdot & \left [ \int d\boldsymbol{x}'\int_0^t dt' \sum_{j=1}^M L_{ij} \left (\boldsymbol{x}-\boldsymbol{x}', t-t' \right ) \times \right. \\
    & \left. \nabla'\frac{\delta H}{\delta \phi_j(\boldsymbol{x}',t')} \right ] + \sqrt{2 T}\nabla\cdot \boldsymbol{\eta}_i(\boldsymbol{x},t).
    \label{modelB}
\end{split}
\end{equation}
Compared to the standard model B~\cite{Bray1994,bray2002theory,hohenberg1977}, we allow here for non-local effects in space and time by considering convolutions of the mobility matrix $\mathbb{L}$ and the thermodynamic driving force $\nabla\delta H/\delta \phi$, which can be understood as the gradient of the chemical potential of the corresponding species. Therefore, $L_{ij}\left (\boldsymbol{x}-\boldsymbol{x}', t-t' \right )$ dictates how the density field of species $i$ at position $\boldsymbol x$ and time $t$ responds to a gradient in the chemical potential of species $j$ at position $\boldsymbol x'$ and time $t'$. Due to thermal fluctuations at temperature $T$, the  fluctuation-dissipation theorem requires the noise correlations to be $\langle \eta_{i\mu}(\boldsymbol{x},t)\eta_{j\nu}(\boldsymbol{x}',t')\rangle = {L}_{ij}(\boldsymbol{x}-\boldsymbol{x}',t-t')\delta_{\mu\nu}$, where $\mu$, $\nu$ are spatial directions; the Kronecker $\delta_{\mu\nu}$ results from spatial isotropy and we have set the Boltzmann constant $k_B=1$. As the kernel $\mathbb{L}$ is a noise correlator, it is  positive semi-definite and symmetric, with the latter property being an example of Onsager's reciprocity relations~\cite{onsager1931}.

Considering small deviations from homogeneous densities for all species, we expand the Hamiltonian (sometimes also referred to as the free energy) $H$ up to quadratic terms in $\phi$~\cite{hohenberg1977,kardar2007, Kruger2017}
\begin{equation}
    H =\frac{1}{2} \sum_{ij}\int d\boldsymbol x \int d\boldsymbol x' \phi_i(\boldsymbol x)\alpha_{ij}(\boldsymbol{x}-\boldsymbol{x}')\phi_j(\boldsymbol x').
    \label{hamiltonianContin}
\end{equation}
$H$ in principle can be found from an expansion of the free energy around the fixed overall species densities
$N_i/V$~\cite{berry2018}. This form includes entropic and energetic effects both from the bulk thermodynamics as well as from interfaces: the former are represented by terms such as $\phi_i\phi_j$, while interfaces between different phases may be accounted for by terms $\nabla\phi_i\nabla\phi_j$ (which can be generated from derivative terms~\cite{Gopinathan10} in $\Balpha$). Spatial homogeneity is encoded in the translational invariance of $\Balpha$. Off-diagonal entries of  $\Balpha$ couple the different components. 

Thermodynamic stability requires the kernel $\Balpha$, which we refer to as the effective interaction, to be positive definite~\cite{Sollich2002,kardar2007}. This may also be understood from the fact that $\Balpha$  is the (functional) Hessian of the free energy~\cite{Weber2019} of the mixture. Closely related to this is the fact that, at equilibrium,  $\Balpha^{-1}\sim \langle \boldsymbol{\phi}\boldsymbol{\phi}^\mathsf{T} \rangle$, the equilibrium (equal time) correlator~\cite{Kruger2017}, i.e.\  $\Balpha^{-1}$ is a correlation kernel.

Using the quadratic form of $H$ in Eq.~\eqref{modelB} yields a closed relation for the mobility $\mathbb{L}$ in terms of time dependent correlation functions. To derive this result, one writes Eq.~\eqref{modelB} in terms of Fourier density modes
\begin{equation}
\phi_i(\boldsymbol{q},t)= \int d\boldsymbol{x}\,e^{-i\boldsymbol{q}\cdot\boldsymbol{x}}\phi_i(\boldsymbol{x},t) = \sum_{k=1}^{N_i}
e^{-i\boldsymbol{q}\cdot\boldsymbol{x}_k(t)}
\end{equation}
as
\begin{eqnarray}
\dot \phi_i(\boldsymbol{q},t) &=& -q^2 
\int_0^t dt'
\sum_{jk} L_{ij}(\boldsymbol{q},t-t')    
\alpha_{jk}(\boldsymbol{q})\phi_k(\boldsymbol{q},t')
\nonumber\\
&&{} + \sqrt{2T} i\boldsymbol{q} \cdot \boldsymbol{\eta}_i(\boldsymbol{q},t)
\label{ModelB_Fourier}
\end{eqnarray}
where $L_{ij}(\boldsymbol{q},\ldots)$, $\alpha_{jk}(\boldsymbol{q})$, $\boldsymbol{\eta}_i(\boldsymbol{q},t)$ are the Fourier transforms of the corresponding quantities in Eq.~\eqref{modelB}.
From this expression one finds directly the equation of motion for the equilibrium structure factor in Fourier space, which is  defined as $\mathbb{S}(\boldsymbol{q},t)=N^{-1}\langle \boldsymbol{\phi}(\boldsymbol{q},t)\boldsymbol{\phi}^\mathsf{T}(-\boldsymbol{q},0)\rangle$. After Laplace transforming in time, this equation becomes in matrix form~\cite{hansen2013}
\begin{equation}
z\mathbb{\hat{S}}-\mathbb{S}_0 = -q^2 \mathbb{\hat{L}}\Balpha\mathbb{\hat{S}}
\end{equation}
where $z$ is the Laplace variable conjugate to $t$ and $\mathbb{S}_0\equiv \mathbb{S}(\boldsymbol{q},t=0)$ is the static (equilibrium, equal-time) structure factor. Using then that~\cite{Kruger2017} $T\Balpha^{-1}/\rho=\mathbb{S}_0$ with $\rho=N/V$ the total density, which incidentally means that  $\Balpha^{-1}$ can be interpreted as a (matrix) thermodynamic compressibility, one can solve for the mobility matrix to obtain in Fourier-Laplace space
\begin{equation}
\hat{\mathbb{{L}}} =
\frac{\rho}{Tq^2} 
(\mathbb{{S}}_0\mathbb{\hat{S}}^{-1}\mathbb{{S}}_0 - z\mathbb{{S}}_0).
\label{mobilitygeneral}
\end{equation}
On the left hand side the mobility is the Fourier-Laplace transform $\mathbb{\hat{L}}\equiv \mathbb{\hat{L}}(\boldsymbol{q},z)$, similarly on the right for the dynamical structure factor
$\mathbb{\hat{S}}\equiv \mathbb{\hat{S}}(\boldsymbol{q},z)$. 
In the following we will focus almost exclusively on the so-called hydrodynamic behavior of the mobility matrix, which is obtained from taking the limits of $z\to 0$ and then $q\to 0$ in Eq.~\eqref{mobilitygeneral}. Equivalently, this approximates the mobility matrix as local in time and space.
For a single component fluid, evaluating the above expression for the mobility is in principle a relatively simple scalar problem, however, doing this quantitatively for e.g.\ systems with slow dynamics remains a challenge~\cite{WisWol14,Stopper15}.
In a mixture with several components, extracting predictions for the full matrix structure is substantially more challenging.

Eq.~\eqref{mobilitygeneral} contains, as such, no fundamentally new information, as it essentially just shifts the problem of determining $\mathbb{L}$ to that of finding the time dependent structure factor $\mathbb{S}(\boldsymbol{q},t)$ and from it $\mathbb{S}_0$ and the Laplace transform $\mathbb{\hat{S}}(\boldsymbol{q},z)$. We will demonstrate in the next subsection, however, that  it can still yield insights into the structure of the mobility matrix. In particular, we will introduce the so-called painted particle model and use this to extract an approximate expression for the full matrix structure of the mobility that is parameterized by a single dimensionless quantity. 

\section{Painted Particle Model}
\label{sec:painted}
\subsection{The model}
In this section we propose a simple model that will allow us to determine the matrix structure of the mobility in mixtures. In this model, a  single-component (or: monodisperse) system is considered, which is then artificially divided into different species. This division can be visualized by coloring particles according to their component affiliation, without modifying their physical properties. We thus introduce $M$ colors, with $N_i$ the number of particles from species $i$, or equivalently with color $i$. For the resulting ``painted particle model'' the mobility matrix $\mathbb{L}(\boldsymbol{q},z)$ in Eq.~\eqref{mobilitygeneral} can be expressed in terms of the structure factor of a {\it single-component fluid}, thereby yielding key insights into the structure and functional form of $\mathbb{L}(q,z)$. The derivation starts from the multi-component time-dependent structure factor for the colored fluid,  
\begin{equation}
\begin{split}
    {S}_{ij}(\boldsymbol{q},t) &= N^{-1}\sum_{k\in \{i\}}\sum_{l\in \{j\}}\langle  e^{-i\boldsymbol{q}\cdot[\boldsymbol{x}_k(t)-\boldsymbol{x}_l(0)]}\rangle\\
    &= c_iS^s(\boldsymbol{q},t)\delta_{ij} + c_ic_j[S(\boldsymbol{q},t)-S^s(\boldsymbol{q},t)].
    \label{ppStructFact}
\end{split}
\end{equation}
Here $S^s(\boldsymbol{q},t)=\langle e^{-i\boldsymbol{q}\cdot[\boldsymbol{x}_1(t)-\boldsymbol{x}_1(0)]}\rangle$ is the self structure factor of the original single-component fluid, and $S(\boldsymbol{q},t)$ is its coherent counterpart; $c_i=N_i/N$ denotes the concentration of species $i$.
The sums over $k$ and $l$ are restricted to particles from the respective species.
Eq.~\eqref{ppStructFact} makes use of the fact that $\langle  e^{-i\boldsymbol{q}\cdot[\boldsymbol{x}_k(t)-\boldsymbol{x}_l(0)]}\rangle$ takes the same value for any pair of particles, as they are all physically identical.

Eq.~\eqref{ppStructFact}  can be stated more compactly in matrix notation, using the concentration vector $\boldsymbol c=(c_1,\dots,c_M)^\mathsf{T}$ and the diagonal matrix $\mathbb{X}$ with components $X_{ij}=c_i\delta _{ij}$:
\begin{equation}
\mathbb{S}(\boldsymbol{q}, t)=\mathbb{X}S^s(\boldsymbol{q},t)+\boldsymbol{cc}^\mathsf{T}[S(\boldsymbol{q},t) -S^s(\boldsymbol{q},t)].\label{Snotation}
\end{equation}
where $\boldsymbol{cc}^\mathsf{T}$ is the outer product of $\boldsymbol c$ with itself. 
For $t=0$, this simplifies to
\begin{equation}
\mathbb{S}_0(\boldsymbol{q})=\mathbb{X}+\boldsymbol{cc}^\mathsf{T}[S_0(\boldsymbol{q})-1]
\label{S0_pp}
\end{equation}
with $S_0(\boldsymbol{q}) \equiv S(\boldsymbol{q},t=0)$. As a side result for later, we give the effective interaction $\Balpha$ for this model, which follows by inverting 
Eq.~\eqref{S0_pp}:
\begin{equation} 
\Balpha_{pp}=\frac{T}{\rho} \mathbb{S}^{-1}_0(\boldsymbol{q}) = \frac{T}{\rho} \left[\left (\frac{1}{S_0(\boldsymbol{q})}-1 \right )\boldsymbol{u u}^\mathsf{T}+\mathbb{X}^{-1}\right].
\label{Ham im terms of S}
\end{equation}
Here $\boldsymbol{u}=(1,\ldots,1)^\mathsf{T}$ is the uniform vector and we have added a subscript to indicate that Eq.~\eqref{Ham im terms of S} holds within the painted particle model. The first term in Eq.~\eqref{Ham im terms of S} is due to interactions and vanishes for an ideal gas~\cite{hansen2013}, while the second term, the ideal gas contribution, results purely from entropy. We see that, in the colored fluid, the effective interaction $\Balpha$ naturally contains off-diagonal terms; these are uniform as every species interacts with every other in the same way.

Returning now to Eq.~\eqref{Snotation}, Laplace transforming this and inserting it alongside Eq.~\eqref{S0_pp} into Eq.~\eqref{mobilitygeneral} yields for the Fourier-Laplace mobility matrix 
\begin{equation}
    \mathbb{\hat{L}}=
    \frac{\rho}{T q^2}\left[(\mathbb{X} - \boldsymbol{cc}^\mathsf{T})\left(\frac{1}{\hat{S}^s}-z\right) + \left(\frac{S_0^2}{\hat{S}} - z S_0\right)\boldsymbol{cc}^\mathsf{T}\right].
    \label{mobilityPainted}
\end{equation}
As for the structure factor matrices above we have abbreviated $S_0\equiv S_0(\boldsymbol{q})$ and $\hat{S}^s\equiv \hat{S}^s(\boldsymbol{q},z)$ here for the scalar, single-component structure factors. We note that the off-diagonal terms of this mobility matrix are of second order in species concentration, which is a necessary condition for preserving positivity of species concentrations at all times~\cite{wahab2011}.

Eq.~\eqref{mobilityPainted} is an important result for this manuscript and the main insight from the painted particle model: it allows us to predict the full mobility matrix from the dynamical structure factor of a single-component fluid. It is notable that this mobility matrix, which is derived from a single-component fluid made up of colored particles, nonetheless carries a nontrivial structure with off-diagonal entries, thus providing a coupling of particles of different colors via the mobility (in addition to the thermodynamic coupling from the effective interaction in Eq.~\eqref{Ham im terms of S}.)    

As explained above, we will mainly focus on the hydrodynamic limit form of the mobility in Eq.~\eqref{mobilityPainted}. Taking first $z\to 0$, it reduces to 
\begin{equation}
\mathbb{\hat{L}}= 
\frac{\rho}{T q^2} 
\frac{1}{\hat{S}^s}[\mathbb{X} - (1-r) \boldsymbol{cc}^\mathsf{T}].
\label{mobilityPainted2}
\end{equation}
Apart from an overall prefactor setting the scale, we see that this mobility matrix depends only on one  dimensionless parameter $r=S_0^2 \hat{S}^s/\hat{S}$ or explicitly
\begin{equation}
r=\frac{S_0^2(\boldsymbol{q}) \hat{S}^s(\boldsymbol{q},z\to 0)}{\hat{S}(\boldsymbol{q},z\to 0)}
\label{r_def}
\end{equation}
This form shows that the matrix structure of the mobility is fully determined by the concentrations and $r$. It also allows to compare our findings to approximations that are typically used for the mobility in the literature~\cite{Mao2019,Shrinivas2021,Vanicat2017,Pagonabarraga2003}. As summarized in the introduction, these are $\mathbb{\hat{L}}\propto \rho\mathbb{X}$ and $\mathbb{\hat{L}}\propto \rho(\mathbb{X}-\rho\boldsymbol{cc}^\mathsf{T})$, and so are contained in Eq.~\eqref{mobilityPainted2} as special cases for the choices $r=1$ and $r=1-\rho$, respectively.

\begin{figure}[hbt]
\begin{center}  
\includegraphics[width=.98\columnwidth]{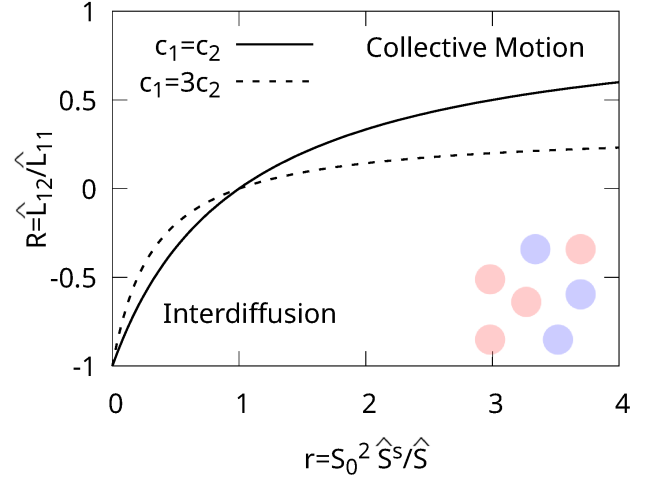}
\caption{Ratio $R$ between off-diagonal and diagonal terms of the mobility indicating the two regimes of interdiffusion ($R<0$) and collective motion ($R>0$). For $r<1$, the fluid can be understood as nearly incompressible, allowing only for exchange in positions of different particles. In the opposite case, $r>1$, an increasing compressibility facilitates collective motion.}
\label{fig:ratio}
\end{center}  
\end{figure}

To obtain further insight into Eq.~\eqref{mobilityPainted2} one can study the ratio between off-diagonal and diagonal elements of the mobility
\begin{equation}
\label{Ratio R}
R_{ij}=\frac{\hat{L}_{ij}}{\hat{L}_{ii}}=\frac{\delta_{ij}-c_j \left(1-r\right)}{1-c_i \left(1-r \right)},
\end{equation}
For $r\to 0$, $R_{ij}$ approaches  $R_{ij}= (\delta_{ij}-c_j)/(1-c_i)$. Physically, this limit can be reached for small $S_0$, i.e.\ a nearly incompressible fluid. Indeed, $R_{ij}$ (for $i\not=j$) is then negative, corresponding to the case of {\it interdiffusion}, whereby different species (colors) diffuse in opposite directions: as the fluid is overall nearly incompressible, species can only exchange positions, while keeping the overall density nearly homogeneous. 
Moving away from the limit of small $r$, the sign of $R_{ij}$ for $i\neq j$ changes at $r=1$, and is positive for $r>1$; the limit of large $r$ can then be interpreted as representing a highly compressible fluid. $R_{ij}>0$ corresponds to the case of {\it collective motion}. Here, different species tend to move in the same direction so that particles of all species diffuse collectively to smoothen inhomogeneities in the total density. Physically, this is possible due to the large compressibility. Fig.~\ref{fig:ratio} illustrates the different regimes for a two-component mixture.

The behavior discussed above has important implications for the demixing of multi-component species. It may also provide a microscopic understanding of the Warren scenario~\cite{Warren1998} outlined above, as the different modes of motion (interdiffusion vs collective diffusion) may dominate at different timescales. We investigate this scenario further in Sec.~\ref{subsec:relaxationTimes}.

\subsection{Relating the painted particle model mobility to physical parameters}
To get more physical insight into the mobility predicted by the painted particle model, we illustrate it for a simple form of the dynamic structure factor. This is motivated by findings from mode-coupling theory~\cite{gotze2009, Janssen2018} for glassy systems. For times large compared to the so-called $\beta$ relaxation time, one has there 
\begin{equation}\label{Structurespecialcase1}
\centering
S(\boldsymbol{q},t)\approx f_q \exp\left[ -\left({t}/{\tau_q}\right)^{\mu_q}\right]S_0(\boldsymbol{q}).
\end{equation}
Here, $\mu_q $ is a stretching exponent, $\tau _q $ is the ($\alpha$-)relaxation time and $0\leq f_q\leq 1$ is the amplitude (or plateau value).
The self-structure factor $S^s(\boldsymbol{q},t)$ is written similarly but with $S_0(\boldsymbol{q})$ replaced by unity; we label the remaining parameters with a superscript $s$.
As explained above we are interested in the modes with small $q$ and small $z$, i.e.\ the regime of large length and timescales, also called the  hydrodynamic limit.
Based on the results of Ref.~\cite{weysser2010structural}, we assume diffusive processes (see Ref.~\cite{Fuchs1999} for a discussion), meaning that in the hydrodynamic limit $\tau_q\approx \lambda^{-2}q^{-2}\tau_0$, with a length $\lambda$, and we use $\mu_q\approx 1$ for the sake of simplicity. With these choices, we find for the mobility tensor (with $f_0$, $\tau_0$ etc.\ indicating the limiting values for $q\to0$) 
\begin{equation}
\lim_{q \to 0} \lim_{z \to 0} \mathbb{\hat{L}}= \frac{\rho\lambda^2}{T}\left [(\mathbb{X}-\boldsymbol{cc}^\mathsf{T})\frac{1}{ \tau^s_0}+\frac{S_0(q\to0)}{f_0 \tau_0} \boldsymbol{cc}^\mathsf{T} \right ].
\label{mobility_explicit}
\end{equation}
To lighten the notation, we use $\mathbb{L}$ without a hat in the following to denote the mobility in the hydrodynamic limit. From Eq.~(\ref{mobility_explicit}), the parameter $r$ introduced in Eq.~\eqref{Ratio R} becomes $r=\tau_0^s S_0(q\to0)/(f_0 \tau_0)$. This illustrates that the transition between interdiffusion and collective motion can arise from a change of compressibility ($S_0(q\to0)$), or from changing the relative values of $\tau_0$ and $\tau_0^s$, i.e.\ by changing the relaxation times for collective and self-diffusion, respectively. Small values of $\tau_0^s$ correspond to rapid, easy self-diffusion and thus favor interdiffusion, while smaller $\tau_0$ lead to a dominance of collective diffusion. 
Of course, as material parameters like density are varied the values of $S_0$, $\tau_0$ and $\tau_0^s$ will all change and these individual effects will combine via the parameter $r$ to determine the dominance of collective diffusion or interdiffusion.

\subsection{Numerical example for mobility: Lattice gas}

\begin{figure}[hbt]
\begin{center}  
\includegraphics[trim={1.0cm 0 1.0cm 0.0cm},clip,width=.49\columnwidth]{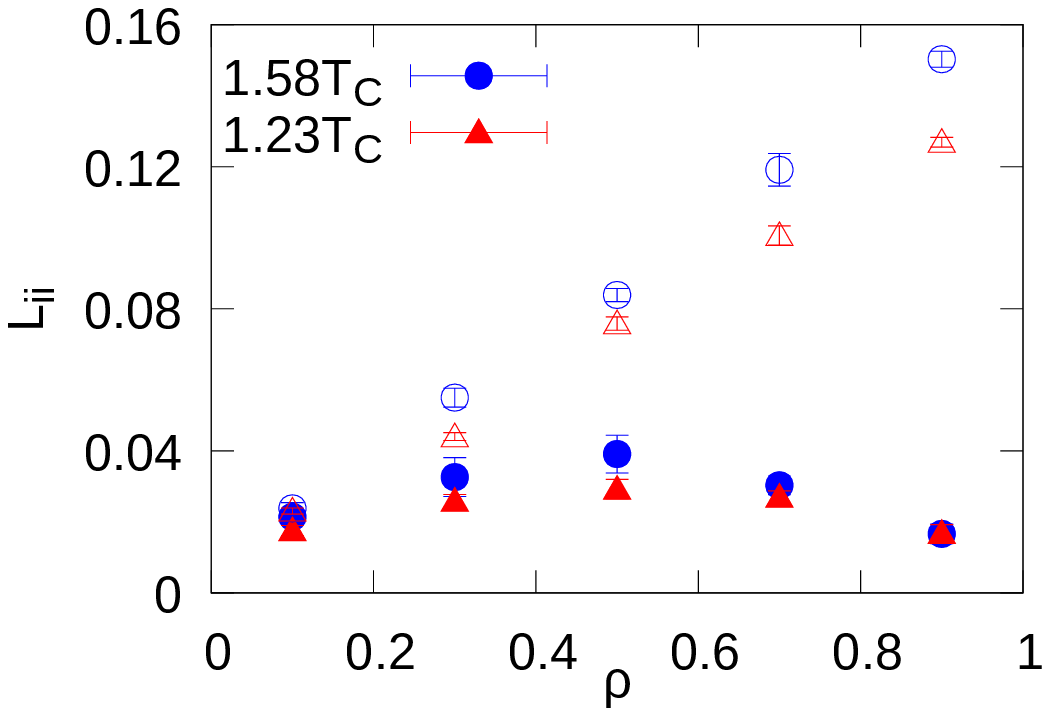}
\includegraphics[trim={1.0cm 0 1.0cm 0.0cm},clip,width=.49\columnwidth]{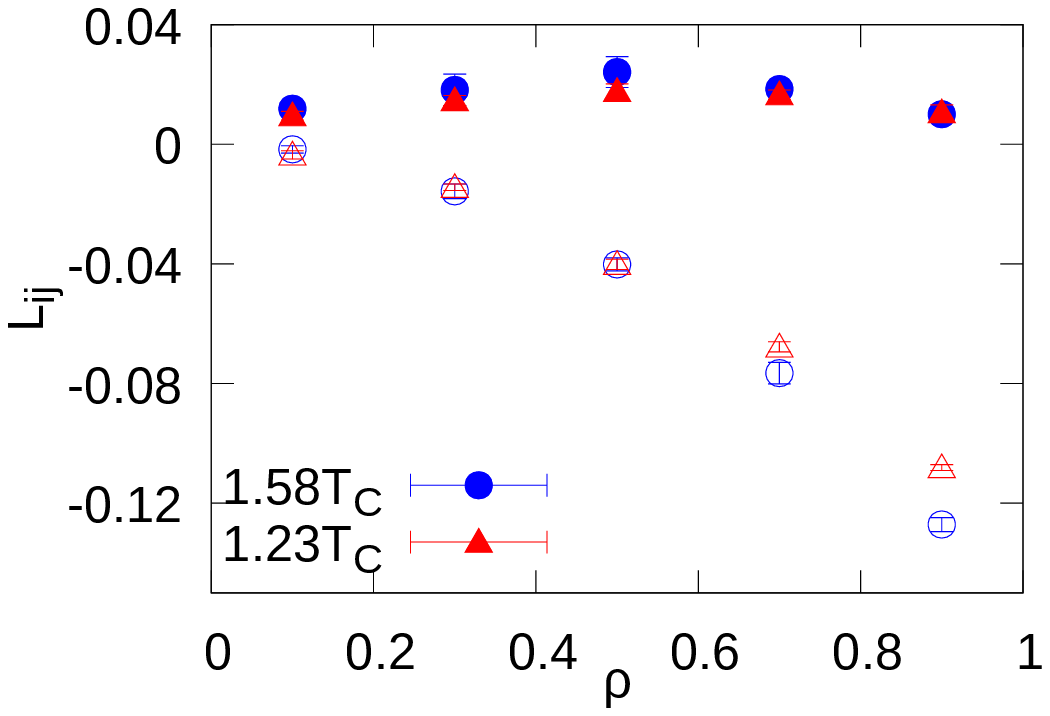}
\caption{Painted particle mobility matrix as a function of total density extracted from Monte Carlo simulations of a single component lattice gas  (two colors with concentrations $c_i=0.5$). (Left) Diagonal elements $L_{11}=L_{22}$, (right) off-diagonal elements $L_{12}=L_{21}$. Shown are data for two different temperatures expressed as multiples of the critical temperature $T_c\simeq 0.568$ obtained from the well-known mapping between the 2d Ising model and the lattice gas. Empty (filled) symbols correspond to the kinetic rule with (without) particle-particle swaps.  Averages of the underlying time-dependent structure factors were taken over $25$ runs, with error bars estimated using the binning analysis~\cite{janke2002} with $5$ blocks.} 
\label{fig:mobiRHO}
\end{center}  
\end{figure}

As a numerical example, we perform Monte Carlo simulations of a two species painted particle lattice gas (i.e., a single component fluid with particles of two different colors) with equal concentrations (see Sec.~\ref{sec:simul} for details of the simulation), and extract $\hat{S}^s$, $\hat{S}$ and $S_0$ (for $z=0$ and the smallest reasonable $q\simeq 0.16$ given our finite system size) for different densities and temperatures. The mobility matrix $\mathbb{L}$ is then obtained via Eq.~\eqref{mobilityPainted}. 
We used two kinetic rules in order to mimic different types of crowding effects, and to investigate their impact on the mobilities. In the first rule, only swaps between particles and vacancies are allowed ($w_{pv}=1$), while swaps between particles are forbidden ($w_{pp}=0$). This rule enhances crowding effects: in high density regions, particles are jammed and cannot diffuse when no vacancies are nearby. The second approach lifts this restriction and allows for particle-particle swaps at the same rate as particle-vacancy swaps ($w_{pp}=w_{pv}=1$). 

Figure~\ref{fig:mobiRHO} shows the resulting mobilities as functions of the total particle density $\rho$, i.e.\ the ratio of total number of particles and number of lattice sites. We first note that the mobilities are only weakly dependent on temperature for the two temperatures investigated. As one changes the density, the two kinetic rules result in very different behaviors. The case with $w_{pp}=0$ (no swaps) produces similar behavior for the diagonal and off-diagonal entries of the mobility: both $L_{ii}$ and $L_{ij}$ approach zero in the limits of low and high densities, with a maximum at $\rho\approx 1/2$. Moreover, since the ratio $L_{ij}/L_{ii}$ is positive for all densities, then according to Fig.~\ref{fig:ratio} the kinetics without particle swaps favours collective motion, and collective density fluctuations will dominate the dynamics. This is consistent with the Warren scenario~\cite{Warren1998}, where stronger crowding is expected to suppress the relative importance of interdiffusion.
On the other hand, the rule with $w_{pp}=1$ (swaps allowed) shows a completely different behavior. Both the absolute values of the diagonal and off-diagonal mobilities increase monotonically with density, and the off-diagonals are now negative. The ratio $L_{ij}/L_{ii}$ is then also negative and the mobilities tell us that interdiffusion of particles of different species is dominant. This is again physically reasonable, given that particle swaps enhance interdiffusion. Finally, for both kinetic scenarios considered, the off-diagonal elements of the mobility matrix are of comparable magnitude to the diagonal ones, emphasizing the need for a full matrix expression for the mobility rather than a diagonal approximation.   
\begin{figure}
\begin{center}  
\includegraphics[width=.98\columnwidth]{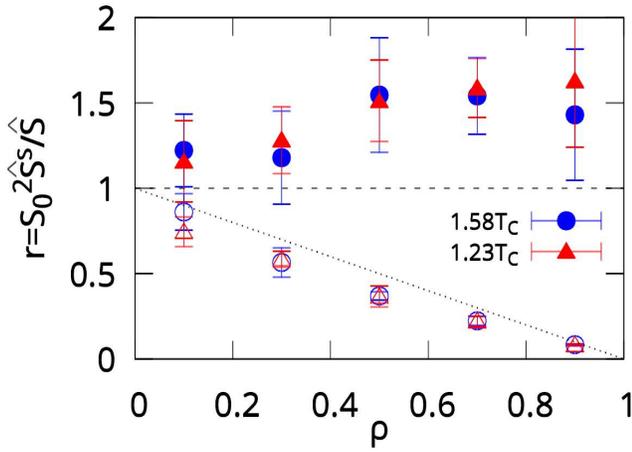}
\caption{Parameter $r=S_0^2\hat{S}^s/\hat{S}$ as a function of total density $\rho$, extracted from the same data as used for Fig.~\ref{fig:mobiRHO}. Filled and empty symbols have the same meaning as there. The dotted and dashed lines show common approximations for the mobility used in the literature~\cite{Mao2019,Shrinivas2021,Vanicat2017,Pagonabarraga2003}, namely $\mathbb{L}\propto \rho\mathbb{X}$ (dashed) and $\mathbb{L}\propto \rho(\mathbb{X}-\rho\boldsymbol{cc}^\mathsf{T})$ (dotted).}
\label{fig:Rrho}
\end{center}  
\end{figure}

We summarize these results in Fig.~\ref{fig:Rrho} by showing the parameter $r=S_0^2\hat{S}^s/\hat{S}$ of Eq.~\eqref{Ratio R} as a function of density for both kinetic rules. 
We observe that the kinetic rule that disallows particle-particle swaps -- and hence has strong crowding effects -- produces a mobility that is not well described by either of the two standard forms of the mobility, $r=1$ or $r=1-\rho$.  The case of allowed particle swaps, on the other hand, is quite well approximated by $r=1-\rho$. This highlights once more the impact of the absence or presence of particle swaps, and more generally of the details of the dynamics, on the mobility matrix. Any successful approximation for $\mathbb{L}$ must then be able to take those details into account.

We expect the mobility we have obtained from the painted particle model to be useful because it is capable of taking into account precisely such system-dependent details of the dynamics,  going beyond existing approximations in the literature based on the strength of crowding effects.

\subsection{Relaxation times}
\label{subsec:relaxationTimes}

\begin{figure}
\begin{center}  
\includegraphics[width=.98\columnwidth]{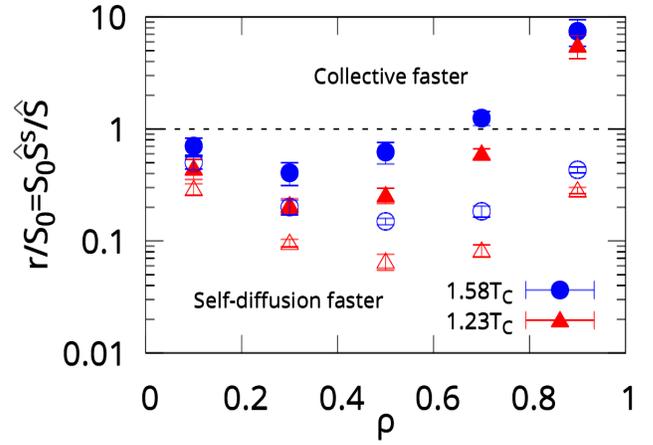}
\caption{Ratio of relaxation timescales for the painted particle model as a function of $\rho$ for two different temperatures, extracted from the same data as used for Fig.~\ref{fig:mobiRHO}. Filled and empty symbols have the same meaning as there. For ratios larger than one (dashed line), collective diffusion is faster while self-diffusion dominates otherwise.
}
\label{fig:timescales}
\end{center}  
\end{figure}

Eq.~\eqref{ModelB_Fourier} shows that in Fourier space, and neglecting non-locality in time of the mobility, the matrix $\mathbb{\Gamma}(q)= q^2  \mathbb{L}(q) \Balpha(q)$ yields the timescales of the system. More precisely, the relaxation times are the inverse eigenvalues of $\mathbb{\Gamma}(q)$. Using the mobility in Eq.~\eqref{mobilityPainted2}, and the painted particle $\Balpha(q)$ in Eq.~\eqref{Ham im terms of S}, we obtain for $\mathbb{\Gamma}$ the form 
\begin{equation}
\label{multiplication of alphaL}
\mathbb{{\Gamma}}_{pp}=\frac{1}{\hat{S}^s}\left[\mathbb{I}-\boldsymbol{c}\boldsymbol{u}^\mathsf{T}
\left(1-\frac{r}{S_0}\right) \right].
\end{equation}
This matrix has two eigenvalues, $\gamma_1=1/ \hat{S}^s(q,z=0)$ and $\gamma_2=S_0/ \hat{S}(q,z=0)$, which correspond to the relaxation rates of the incoherent and coherent correlations, respectively. The ratio between them is $\gamma_2/\gamma_1=r/S_0=S_0 \hat{S}^s/\hat{S}$. Fig.~\ref{fig:timescales} shows this quantity as extracted from Monte Carlo simulations, again for the numerically estimated limit $q\to 0$.
Here, a clear competition between thermodynamics, represented by $\Balpha$, and kinetics arises. While the mobility $\mathbb{L}$ in Fig.~\ref{fig:mobiRHO} predicts only collective motion for $w_{pp}=0$, i.e.\ $r>1$, interdiffusion has the shorter timescale $1/\gamma_1=\hat{S}^s$ at intermediate densities, as can be seen from $\gamma_2/\gamma_1<1$, and therefore dominates the relaxation. This is a direct consequence of the interplay of the thermodynamic $\Balpha$ and the purely kinetic $\mathbb{L}$. In spite of the mobility driving towards collective motion, the tendency in $\Balpha$ to create interdiffusion due to entropy prevails at short times. At high densities, crowding effects introduced by forbidding particle-particle swaps strongly suppress interdiffusion as compared to collective motion as can be seen by the filled symbols in Fig.~\ref{fig:mobiRHO} crossing the dashed line at $\gamma_2/\gamma_1=1$. This behavior is directly connected to the Warren scenario~\cite{Warren1998} as we discussed previously. Overall, the painted particle model, as introduced above, not only provides a mobility that can represent both collective motion and interdiffusion, but also accounts for the interesting interplay between this mobility and thermodynamic effects.

\subsection{(Approximate) determination of the mobility matrix in simulations or experiments}
\label{determinationMobility}
\begin{figure}[hbt]
  \begin{center}  
    \includegraphics[width=.99\columnwidth]{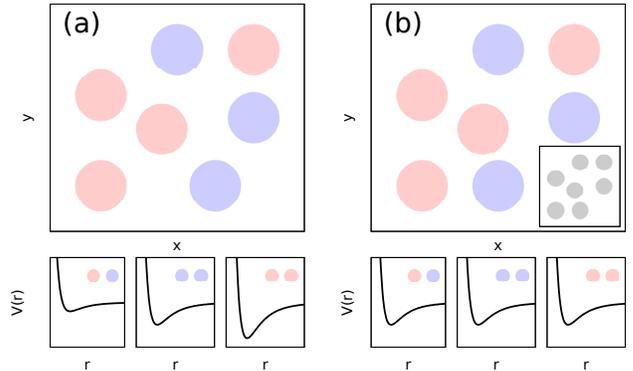}
    \caption{The mobilities of a multicomponent mixture with (a) distinguishable interactions can be approximated by (b) a painted particle model with indistinguishable interactions, which is equivalent to (inset) a single component fluid.}
    \label{fig:paintedParticleSketch}
  \end{center}  
\end{figure}

How can the mobility matrix be found in a true multi-component system? In principle, the mobility in Eq.~\eqref{mobilitygeneral} can be determined in simulations or experiments from the full dynamical structure factor matrix. This, however, will typically be impractical, especially once many components are present in the mixture. We thus provide two levels of approximation, using  Eq.~\eqref{mobilityPainted} (see also Fig.~\ref{fig:paintedParticleSketch}). 

The {\it painted particle approximation} relies on the assumption that Eq.~\eqref{mobilityPainted} remains valid in systems with physically different species.  Using it, the scalars $\hat S$, $\hat{S}^s$ and $S_0$ may be evaluated in different ways. 

In a first method, one measures the scalars $\hat{S}$, $\hat{S}^s$ and $S_0$ by summing over all species components, hence ignoring the fact that they are fundamentally different. This approach allows for the structure factors to contain information on typical timescales resulting from the distinguishable (species-dependent) interactions, while assuming such timescales to be common to all species. 

A second method measures the scalars $\hat{S}$, $\hat{S}^s$ and $S_0$ in a single component fluid that is close to the mixture under consideration, see Fig.~\ref{fig:paintedParticleSketch}. This then in turn requires finding a suitable single-component fluid that closely resembles the mixture. Methods for this will be discussed in detail in Sec.~\ref{sec:simul} below.  

\section{Dynamics after a quench}
 \label{sec:quench}

As an application of the mobility matrix we have derived from the painted particle model, we will analyze in this section the spatial correlations after a temperature quench, within a linear (Gaussian) theory. This will make contact with previous work~\cite{Rohwer2017}, and will give analytical insights into the impact of the mobility matrix for quenched mixtures. The results will be compared to simulations in Sec.~\ref{sec:simul}.

\subsection{General case with time-local mobility}
The mobility~\eqref{mobilityPainted} encodes the kinetics of a mixture of $M$ components in equilibrium. In our discussion of the quench scenario we assume that these expressions also hold out of equilibrium. We expect this to be the case for mild quenches, where the imposed temperature change is small and no phase coexistence boundaries are crossed.

We further employ the approximation that the mobility is local in time. This will allows us to derive the explicit time dependence of the correlations. We start from the Fourier mode time evolution~\eqref{ModelB_Fourier} and write this as
\begin{equation}
\dot{\phi}_i(\boldsymbol{q},t)=-\sum_j\Gamma_{ij}(\boldsymbol{q}) \phi_j(\boldsymbol{q},t)+\sqrt{2T}\tilde{\eta}_i(\boldsymbol{q},t).
\label{1.11}
\end{equation}
We have used (as before)  $\mathbb{\Gamma}(\boldsymbol{q})= q^2  \mathbb{L}(\boldsymbol{q}) \Balpha(\boldsymbol{q})$ and introduce the noise $\tilde{{\eta}}_i=\boldsymbol\nabla\cdot \boldsymbol{\eta}_i$, with $\langle \tilde\eta _{i\mu}(\boldsymbol{q},t)\tilde\eta_{j\nu}(\boldsymbol{q}',t')\rangle =q^2V L_{ij}(\boldsymbol{q}) \delta_{\mu\nu} \delta(t-t')\delta_{\boldsymbol{q},-\boldsymbol{q}'} $. The solution to this equation reads, now using vector notation for the field vector $\boldsymbol{\phi}$,
\begin{equation} 
\boldsymbol{\phi}(\boldsymbol{q},t)=e^{-\mathbb{\Gamma}t}\boldsymbol{\phi}(\boldsymbol{q},0)+\sqrt{2T}\int_0^t{d \tau\, e^{-\mathbb{\Gamma}(t-\tau)} \tilde{\boldsymbol{\eta}}(\boldsymbol{q},\tau)}.
\label{1.1sloution}
\end{equation}
Eq.~\eqref{1.1sloution} is valid for any initial condition, with the other parameters in the solution then reflecting the situation after the quench at $t=0$. We now consider a quench from an initial temperature $T_i$ to a final temperature $T$~\cite{Gopinathan10} and compute the time evolution of the correlation matrix in Fourier space, defined as 
$\mathbb{C}(\boldsymbol{q},t)=V^{-1}\langle \boldsymbol\phi(\boldsymbol{q},t) \boldsymbol\phi^\mathsf{T}(-\boldsymbol{q},t) \rangle$. The difference between the correlations at time $t$ and the initial correlations, $\Delta \mathbb{C}\equiv\mathbb{C}(\boldsymbol{q},t)-\mathbb{C}(\boldsymbol{q},t=0)$ is then found to be
\begin{equation}
\label{Delta C in Fourier space}
\Delta \mathbb{C}=-\mathbb{\Delta}+e^{-\mathbb{\Gamma}t}\mathbb{\Delta} e^{-\mathbb{\Gamma}^\mathsf{T} t},
\end{equation}
where we have set $\mathbb{\Delta}=T_i\Balpha_i^{-1}-T\Balpha^{-1}$ with the shorthand $\Balpha_i\equiv\Balpha(T_i)$. We have also used the fact that before the quench the system is -- by assumption -- equilibrated at temperature $T_i$, which gives $V^{-1}\langle \boldsymbol{\phi}(\boldsymbol{q},t=0)\boldsymbol{\phi}^\mathsf{T}(-\boldsymbol{q},t=0)\rangle=T_i{\Balpha}_i^{-1}$. 

\subsection{Locality in time and space: hydrodynamic Limit} 

Eq.~\eqref{Delta C in Fourier space} is valid for any $\boldsymbol{q}$-dependent $\Balpha$ and $\mathbb{L}$. Performing the inverse Fourier transform, however, to extract real-space information depends on the functional form of these matrices. A further simplification can be made if we assume that $\Balpha$ does not depend on $q$, which corresponds to a local interaction in the Hamiltonian~\eqref{hamiltonianContin}. We will also assume that $\mathbb{L}$ is approximately independent of $q$. Performing an eigen-expansion of the matrices in~\eqref{Delta C in Fourier space}, the inverse Fourier transform can then be carried out and we obtain the correlation matrix in real space
\begin{equation}\label{correlation in real space}
\Delta\mathbb{C}(X,t)=
\sum_{ik}\frac{e^{-\frac{X^2}{4(\gamma_i+\gamma_k)t}}}{[2t(\gamma_i+\gamma_k)]^{\frac{d}{2}}} \boldsymbol l_i^\mathsf{T}\mathbb{\Delta}\boldsymbol{l}_k \boldsymbol{r}_i \boldsymbol{r}_k^\mathsf{T}
\end{equation}
in terms of the left and right eigenvectors $\boldsymbol{l}_i$ and $\boldsymbol{r}_i$ of $\mathbb{\Gamma}$, respectively; the $\gamma_i$ are the corresponding eigenvalues~\cite{Risken}. The spatial distance of points in the correlator is $X$, and $d$ is the spatial dimension.

The double sum over eigenvalues  in the last expression illustrates the richness of behaviors in multi-component mixtures: there are in general $M(M+1)/2$ distinct $\gamma_i + \gamma_k$ and hence distinct exponentials in~\eqref{correlation in real space}. As the time $t$ since the quench varies, different timescales will dominate until, finally, the correlation decays as a power law in time, $t^{-d/2}$, as is the case for the single component fluid~\cite{Rohwer2017}.

\subsection{Analytical example: Ideal gas with mobility matrix}
\label{L-arbitrary}
In this subsection, we aim to provide more analytical insights into the dynamics after quench, and to highlight the role of the matrix structure of the mobility. We assume that $\Balpha=\alpha(T)\mathbb{I}$ is the identity with prefactor  $\alpha$, a function of temperature. This may represent a gas of ideal particles with each species having equal density $c_i$, so that $\mathbb{X}$ is proportional to the identity matrix; from Eq.~\eqref{Ham im terms of S},  $\Balpha=\alpha(T)\mathbb{I}$ then follows in the absence of interactions \footnote{Strictly, for an ideal gas, $T/\alpha(T)$ is independent of temperature, so that a quench has no effect. We thus assume that the initial distribution of particles can be manipulated by some other means.}. In contrast, we let $\mathbb{L}$ be an arbitrary (symmetric and positive)  matrix. This allows us to obtain an explicit expression for the time evolution of the correlation matrix in real space after taking the inverse Fourier transform, and to make a connection to earlier studies on single component fluids~\cite{Kruger2017}. 

With the above choices, $\mathbb{\Gamma}= q^2  \mathbb{L} \Balpha$ is symmetric and $\mathbb{\Delta}$ is a multiple of the identity matrix, so that the second term in Eq.~\eqref{Delta C in Fourier space} simplifies to~\cite{Gopinathan10}
\begin{equation}
    e^{-\mathbb{\Gamma}t}\mathbb{\Delta} e^{-\mathbb{\Gamma}^\mathsf{T} t}=\mathbb{\Delta} e^{-2\mathbb{\Gamma} t}.
\end{equation}
This yields the correlation matrix (change)
\begin{equation}\label{eq:C}
\centering
\Delta\mathbb{C}(\boldsymbol{q},t)=\left (\frac{T_i}{\alpha_i}-\frac{T}{\alpha}\right )\left(e^{-2q^2 \alpha \mathbb{L}t}-\mathbb{I}\right)
\end{equation}
where $\alpha_i$ and $\alpha$ are the scalar prefactors $\alpha(T)$ at the initial and final temperatures, respectively. Focusing on the case of $d=2$ spatial dimensions (as in our simulations in Sec.~\ref{sec:simul}) we  take the inverse Fourier transform to obtain
\begin{equation}\label{correlation in real space for small tem difference}
\centering
\Delta\mathbb{C}(X,t)=\left (\frac{T_i}{\alpha_i}-\frac{T}{\alpha}\right )\frac{1}{8\pi \alpha t}\mathbb{L}^{-1} e^{-\frac{X^2\mathbb{L}^{-1}}{8\alpha t}}.
\end{equation}
where we have ignored the $\delta$ function at the origin resulting from the last term in Eq.~\eqref{eq:C}.
Recalling that $\mathbb{L}$ is symmetric and positive definite, we may again invoke an expansion in terms of normalized eigenvectors $\boldsymbol{l}_i$ and eigenvalues $L_i$ to obtain
\begin{equation}\label{alpha is arbitarry and L is identity }
\Delta\mathbb{C}(X,t)=\left (\frac{T_i}{\alpha_i}-\frac{T}{\alpha}\right )\frac{1}{8\pi\alpha t} \sum_i\frac{1}{L_i}e^{-\frac{X^2}{8\alpha t L_i  }} \boldsymbol{l}_i \boldsymbol{l}_i^\mathsf{T}.
 \end{equation}
For the specific case of a two-component mixture, $M=2$, $\mathbb{L}$ has two eigenvalues $L_1$ and $L_2$ and the eigenvector outer products $\boldsymbol{l}_i \boldsymbol{l}_i^\mathsf{T}$ can be written in the form
 \begin{align}
    \boldsymbol{l}_i \boldsymbol{l}_i^\mathsf{T}=\frac{1}{a_i^2+1}\begin{pmatrix}
a_i^2 & a_i \\
a_i & 1\\
\end{pmatrix}.
 \end{align}
As both diagonal entries of this matrix are non-negative,  the sign of the diagonal entries of $\Delta\mathbb{C}$ is independent of time and given by the sign of $\left (\frac{T_i}{\alpha_i}-\frac{T}{\alpha}\right )$. In contrast, one can easily show that $a_1$ and $a_2$ have opposite signs, so that the off diagonal elements of $\Delta\mathbb{C}$ can be either positive or negative, and change sign as a function of time $t$. Using $a_1/(a_1^2+1)+a_2/(a_2^2+1)=0$,  which follows from orthogonality of the eigenvectors, we may write the single off-diagonal element as
 \begin{align}\label{correlation c12 for case1}
\Delta C_{12}(X,t)&=\left(\frac{T_i}{\alpha_i}-\frac{T}{\alpha} \right)\frac{a_2}{ 8 \pi \alpha t(a_2^2+1)}\times\notag\\ &\left[\frac{e^{-\frac{X^2}{8 \alpha L_2 t}}}{L_2}-\frac{e^{-\frac{X^2}{8 \alpha L_1 t}}}{L_1}\right].
\end{align}
We can assume without loss of generality that $L_2>L_1$. Thus, for sufficiently short times  (see below for details on the timescale), where the exponential functions rapidly go to zero,
 \begin{align}
     \frac{e^{-\frac{X^2}{8 \alpha L_2 t}}}{L_2}> \frac{e^{-\frac{X^2}{8 \alpha L_1 t}}}{L_1}.
 \end{align}
For large times, on the other hand, the exponential functions approach unity and
\begin{align}
     \frac{e^{-\frac{X^2}{8 \alpha L_2 t}}}{L_2}< \frac{e^{-\frac{X^2}{8 \alpha L_1 t}}}{L_1}.
 \end{align}
 The off-diagonal element $\Delta C_{12}$ thus shows a change of sign  at some $t_s$ where the two terms are equal, given explicitly by
\begin{equation}
   t_s=\frac{X^2(L_2-L_1)}{8\alpha L_1 L_2 \ln{\frac{L_2}{L_1}}   }.
\end{equation}
Overall, depending on the sign of $a_2 \left(\frac{T_i}{\alpha_i}-\frac{T}{\alpha} \right)$, the off-diagonal elements of the correlator change sign in time from positive to negative or vice versa. 

\begin{figure}[hbt]
  \begin{center}  
    \includegraphics[width=.98\columnwidth]{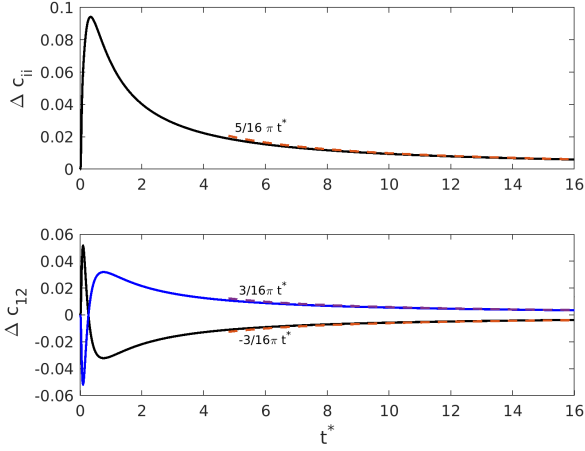}
    \caption{Different entries of the rescaled correlation matrix $\Delta c_{ij}= \Delta C_{ij} X^2/(\frac{T_i}{\alpha_i}-\frac{T}{\alpha})$ versus dimensionless time $t^{*} =\alpha L_{2}t/X^2$. Top: Diagonal elements $\Delta c_{11}=\Delta c_{22}$, bottom: off-diagonal element $\Delta c_{12}$.
    Here we cover both cases  ${L}_{11}/L_2={L}_{22}/L_2=5/8$, ${L}_{12}/L_2=\pm 3/8$ and $\Balpha=\alpha \mathbb{I}$. As discussed in the text, the diagonal terms are positive for all times while the off-diagonals change from positive to negative ($L_{12}/L_2=3/8$, black) or vice-versa ($L_{ij}/L_2=-3/8$, blue). The dashed lines show the asymptotic behavior for large $t^*$.
    }
    \label{fig1}
  \end{center}  
\end{figure}

Fig.~\ref{fig1} shows an example case with  $L_{11}/L_2=L_{22}/L_2=5/8$ and  $L_{12}/L_2=\pm 3/8$. The diagonal elements of $\Delta\mathbb{C}$ are identical, and independent of the sign of $L_{12}$. The sign of the off-diagonal part depends on the sign of $L_{12}$: positive $L_{12}$ (black curve in Fig.~\ref{fig1}) yields positive $a_2$ and negative $L_{12}$ gives negative $a_2$ (blue curve). For the figure we have introduced a dimensionless time, $t^{*} =\alpha L_{2}t/X^2$, to write Eq.~\eqref{correlation c12 for case1} as 
\begin{equation}\label{correlation c12 for case2}
\frac{\Delta C_{12}(X,t)X^2}{\frac{T_i}{\alpha_i}-\frac{T}{\alpha} }=\frac{ a_2}{ 8 \pi (a_2^2+1)t^*} e^{-\frac{1}{8  t^*}}\left[1-\frac{L_2}{L_1}e^{-\frac{(L_2-L_1)}{8 L_1 t^*}}\right],
 \end{equation}
and similarly for the diagonal elements. The right hand side of Eq.~\eqref{correlation c12 for case2} is the curve shown in Fig.~\ref{fig1}.

\begin{figure}[hbt]
  \begin{center}  
    \includegraphics[width=.98\columnwidth]{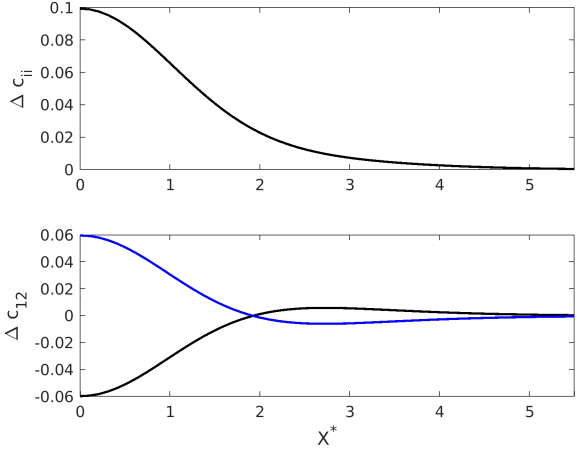}
    \caption{Different entries of the rescaled correlation matrix $\Delta c_{ij}=\Delta C_{ij}\alpha L_2 t/( \frac{T_i}{\alpha_i}-\frac{T}{\alpha}) $ versus dimensionless position $X^*{}^2=X^2/\alpha L_2 t$. Parameters as in Fig.~\ref{fig1}. As there, the diagonal terms (top) are positive while the off-diagonal (bottom) changes from negative to positive ($L_{ij}/L_2=3/8$, black) or vice-versa ($L_{ij}/L_2=-3/8$, blue). 
    }
    \label{fig2}
  \end{center}  
\end{figure}

Fig.~\ref{fig2} shows the correlations after a quench as a function of spatial separation $X$. Here, we introduce the dimensionless distance $X^*{}^2=X^2/\alpha L_2 t\equiv 1/t^*$ to rewrite  Eq.~\eqref{correlation c12 for case1} as 
\begin{equation}
 \frac{\Delta C_{12}\alpha L_2 t}{ \frac{T_i}{\alpha_i}-\frac{T}{\alpha} }=\frac{a_2}{ 8 \pi (a_2^2+1)}  e^{-\frac{X^*{}^2}{8}}\left[1-\frac{L_2}{L_1}e^{-\frac{X^*{}^2(L_2-L_1)}{8 L_1}}\right],\label{scaledX}
 \end{equation}
with again similar expressions for the diagonal elements.

This example illustrates the importance of the mobility matrix: even though the mixture components are uncoupled thermodynamically ($\Balpha$ is diagonal), the mobility causes correlations between the components at intermediate times. Here, the case of $L_{12}$ positive corresponds to $R>0$ in Eq.~\eqref{Ratio R}, i.e.\ collective diffusion. Indeed, for short times, the sign of $\Delta C_{12}$ follows the sign of $\Delta C_{ii}$, indicating collective diffusion. $\Delta C_{12}$ then changes sign as a function of $t$, and has the opposite sign compared to $\Delta C_{ii}$ at large times. This shows that the interplay of $\mathbb{L}$ and $\mathbb{\alpha}$, as time evolves, can lead to a more complex behavior than anticipated in Fig.~\ref{fig:ratio}, where we focused only on the dominant (fastest) relaxation process. The anti-correlation ($\Delta C_{12}<0$) at large times may be interpreted as an onset of demixing, an interpretation which may also be seen in Fig.~\ref{fig2}: As a function of distance $X$, the off-diagonal elements change sign, indicating a transient structure in the mixture. This shows that the mobility can have a strong influence on the dynamics, and possibly  intermediate phases. Using negative $L_{12}$ reverts this discussion. It corresponds to $R<0$, and here, the short times are dominated by interdiffusion.

\begin{figure}[hbt]
  \begin{center}  
    \includegraphics[width=\columnwidth]{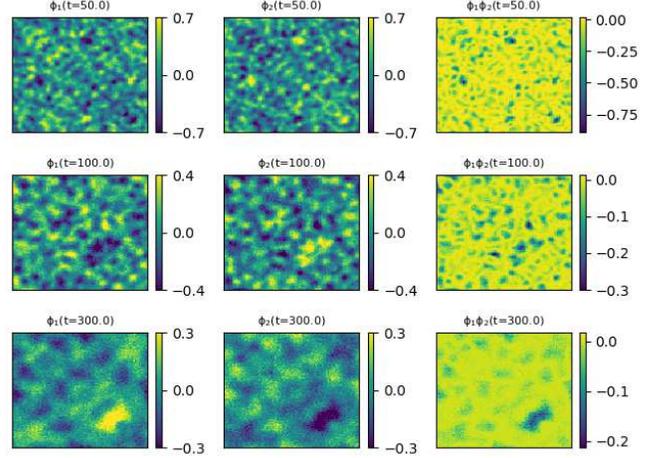}
    \caption{Typical configurations in real space obtained by sampling the Fourier modes according to Eq.~\eqref{eq:C}. The colors represent values of $\phi_1$ (left), $\phi_2$ (center) and $\phi_1\phi_2$ (right) in a binary mixture. For the snapshots shown we chose $\alpha_i=\alpha=1$, $T_i=4$, and $T=2.5$. The box has size $L_{\textrm{box}}=100$ and the mobility parameters are $L_{11}=L_{22}=5$ and $L_{12}=-3$.}
    \label{figSnaps}
  \end{center}  
\end{figure}

For physical insight it is helpful to visualize the above discussion further. We illustrate the structures after a quench by showing typical configuration snapshots at different times, sticking as before to a diagonal form of $\Balpha=\alpha\mathbb{I}$. We sample the Fourier modes from a normal distribution with covariance matrix given by Eq.~\eqref{eq:C} and perform the inverse transform to obtain the contour graphs in Fig.~\ref{figSnaps}. 

At early times (top row in Fig.~\ref{figSnaps}), the graphs show fluctuations and correlations at small length scales, as expected from Figs.~\ref{fig1} and \ref{fig2}. With increasing time (center and bottom rows) the correlated regions in space grow and the amplitude of the fluctuations decreases, again both as expected. Notably, a clear correlation exists between the two species for all times shown. This is visible from the last column, as well as from a comparison of the left and middle columns: $\phi_1$ is large (small) in regions where $\phi_2$ is small (large). We re-emphasize that this onset of demixing is driven purely kinetically here as $\Balpha$ is diagonal. If the mobility matrix were diagonal, a vanishing spatial average of $\phi_1\phi_2$ would result (not shown). 

\subsection{Quench in a colored single component fluid}\label{scf}

As a second special case of our general calculations for correlations after a quench, we consider next a quench from initial temperature $T_i$ to final temperature $T$ governed by painted particle dynamics, i.e.\ by $\mathbb{L}$ and $\Balpha$ from Eqs.~\eqref{mobilityPainted} and~\eqref{Ham im terms of S}, respectively. This will also connect with previous studies~\cite{Rohwer2017,Rohwer2018} for single component fluids. From the inverse of \eqref{Ham im terms of S},
$T\Balpha_{pp}^{-1}(T)=\rho [\mathbb{X} - \boldsymbol{cc}^\mathsf{T}(S_0 - 1)]$, it follows, due to cancellation of the ideal gas term, that $\mathbb{\Delta}=\rho(S_0^{(i)}-S_0)\boldsymbol{cc}^\mathsf{T}$ and
\begin{equation}
\Delta\mathbb{C}(q,t)=  -\rho (S_0^{(i)}-S_0)\left [\boldsymbol{cc}^\mathsf{T} - e^{-\mathbb{\Gamma}t}\boldsymbol{cc}^\mathsf{T} e^{-\mathbb{\Gamma}^\mathsf{T} t}\right ],
 \label{correlation in q space}   
 \end{equation}
In these expressions, $S_0$ is the static structure factor at the final temperature as before, while $S_0^{(i)}$ is the one at $T_i$. 

Expanding the exponential functions and using that in the present scenario $\mathbb{\Gamma}$ is given by Eq.~\eqref{multiplication of alphaL}, terms of the form $\boldsymbol{c}\boldsymbol{u}^ \mathsf{T}\dots \boldsymbol{c}\boldsymbol{u}^ \mathsf{T} \boldsymbol{c}\boldsymbol{c}^\mathsf{T}\boldsymbol{u}\boldsymbol{c}^ \mathsf{T}\dots \boldsymbol{u}\boldsymbol{c}^ \mathsf{T}$ appear, which simplify to $\boldsymbol{c}\boldsymbol{c}^\mathsf{T}$ because $\boldsymbol{c}^\mathsf{T}\boldsymbol{u}=\boldsymbol{u}^ \mathsf{T}\boldsymbol{c}=1$.
The correlator is thus found as 
\begin{equation}
\Delta \mathbb{C}(q,t)=\rho(S_0^{(i)}-S_0)\left(1-e^{-2S_0t/\hat{S}}\right)\boldsymbol{cc}^\mathsf{T}. 
\end{equation}
It is a rank one matrix, as expected from the equivalence of different colors, and all elements of the correlation matrix have the same time dependence. The time dependence is governed by a single timescale, namely $\hat{S}/S_0$, again as one would expect for a single component fluid. Interestingly, the timescale associated with self-diffusion in the dynamical matrix~\eqref{multiplication of alphaL} plays no role in the quench dynamics and is absent from the final expression.

The single component fluid~\cite{krugerdean2017b} can finally be easily recovered by summing over all elements of $\Delta\mathbb{C}$; explicitly one has
\begin{equation}
\boldsymbol{u}^\mathsf{T}\Delta \mathbb{C}\boldsymbol{u}=\rho(S_0^{(i)}-S_0)(1-e^{-2S_0t/\hat{S}})
\end{equation}
and this is independent of $\boldsymbol{c}$ as it must be.
A useful limiting case is given by infinite initial temperature and will become important later when we compare our predictions to numerical simulations of an interacting lattice gas with volume exclusion. For a fluid with finite energy barriers, one then has the ideal gas form $S_0^{(i)}=S_0(T\to\infty)=1$. In case of a hard core interaction, on the other hand, as e.g.\ in a lattice gas,  $S_0^{(i)}=S_0(T\to\infty)\simeq 1-\rho$. 

In order to describe nontrivial physics, we have to go beyond a single component fluid. A true multi-component mixture can then be approximately described by using the mobility of the reference single component fluid together with the appropriate form of $\Balpha$, as described above in Sec.~\ref{determinationMobility}. The difference between $\Balpha$ and the painted particle version $\Balpha_{pp}$ then gives rise to additional timescales, as will be discussed in the next Section and in the context of the simulations in Sec.~\ref{sec:simul} below.

\subsection{Correction to painted particle model}

Writing the thermodynamic and kinetic parameters for a generic multi-component mixture with the painted particle model as a baseline, we have $\Balpha=\Balpha_{pp}+\tilde{\Balpha}$ and $\mathbb{L}=\mathbb{L}_{pp}+ \tilde{\mathbb{L}}$. We remain with the assumption that the painted particle description is appropriate for the kinetics and so neglect $\tilde{\mathbb{L}}$. This yields  $\mathbb{\Gamma}=\mathbb{\Gamma}_{pp}+\mathbb{L}_{pp}\tilde{\Balpha}$. Furthermore it is useful to assume that  $\tilde{\Balpha}$ is independent of temperature, which is true if it is a second virial term~\cite{hansen2013}. The relaxation timescales are then given by the inverse of the eigenvalues of
\begin{equation}
    \mathbb{\Gamma} = \mathbb{\Gamma}_{pp} + \frac{\rho}{T \hat{S}^s}\left [ \mathbb{X} - (1-r)\boldsymbol{cc}^\mathsf{T} \right ]\tilde{\mathbb{\Balpha}}.
    \label{gammaDistinguishable}
\end{equation}
The second term will then yield further timescales, beyond those found in Sec.~\ref{scf}. A tractable example that has been frequently studied~\cite{castro2018,carugno2022} is $\tilde{\Balpha}=\boldsymbol{\sigma\sigma}^\mathsf{T}$ where $\boldsymbol{\sigma}=\{ \sigma_i \}$ is the vector of a single property $\sigma_i$ of species $i$, e.g.\ its size or charge, that distinguishes it from other components. The second term in~\eqref{gammaDistinguishable} is then of rank one and adds a single new timescale to the evolution of the system.

For general cases where $\tilde{\Balpha}$ is dependent on temperature, the analog Eq.~\eqref{gammaDistinguishable} becomes more complicated. In practical applications in Sec.~\ref{sec:simul} we therefore find $\Balpha$ numerically and use it directly in Eq.~\eqref{Delta C in Fourier space}. 

\section{Simulations}
\label{sec:simul}
We consider a multi-component lattice gas where each site on a square lattice is occupied by either a single particle or a vacancy. The Hamiltonian of a configuration is given by
\begin{equation}
    H_d = -\frac{1}{2}\sum_{s_k} \sum_{s_l\in\partial s_k} \sum_{i,j=1}^M n_i(s_k) n_j(s_l) \epsilon_{ij}
\end{equation}
where $n_i(s_k)$ is the occupancy number of site $s_k$ by species $i$, which is equal to one if site $s_k$ is occupied by a particle of species $i$ and zero otherwise. The sum over $s_l$ runs over nearest neighbors of $s_k$ and the energy cost of a bond between species $i$ and $j$ is given by $\epsilon_{ij}$. The above Hamiltonian is the lattice equivalent of Eq.~\eqref{hamiltonianContin} where only nearest neighbors interactions are considered as indicated by the restriction $s_l\in \partial s_k$.

We evolve the system towards equilibrium using kinetic Monte Carlo, specifically Kawasaki dynamics with Glauber acceptance probability $w(1 + e^{\beta \Delta H_d})^{-1}$. Here $\Delta H_d$ is the energy change in going from the current to the proposed configuration. The only movements allowed are swaps between neighboring sites with rate $w=w_{pp}$ if both sites are occupied by particles and with rate $w=w_{pv}$ if one of the sites is occupied by a vacancy. The two different rates allow us to control for crowding effects at high densities. This kinetic approach, although performed with discrete time dynamics, is known to provide a good approximation to the dynamics defined directly in continuous time with corresponding transition rates~\cite{Bal2014}.

First, given the general result for the painted particle mobility in Eq.~\eqref{mobilityPainted}, we extract equilibrium values of $\hat{S}$, $\hat{S}^s$ and $S_0$ for a single-species reference system, averaged over $25$ different runs and evaluate $\mathbb{L}$. This is done by discarding the initial $10^5$ Monte Carlo steps and using the data of the subsequent $10^5$ steps (or until the structure factors decay to zero for the $q$ of interest). Since we are interested in the hydrodynamic limit, we focus on small values of $q\simeq 0.16$ and $z=0$. 
Following the discussion in Sec.~\ref{determinationMobility},
we need to define the reference fluid via a typical interaction $\epsilon$ that allows us to recover the mobility of a multi-species system with distinguishable interactions. In the following examples, we always do this by setting the interaction in the reference fluid to $\epsilon=(1/M)^2\sum_{ij}\epsilon_{ij}$. 

Our mixture of interest will have three species and distinguishable, i.e.\ species-dependent interactions. We perform simulations of this system for a quench and measure the time evolution of the correlation matrix until it reaches the steady state. From the relationship $T\Balpha^{-1}=\rho\mathbb{S}_0$ in equilibrium we extract $\Balpha$ and insert it, together with the mobility $\mathbb{L}$ obtained from the reference fluid, into~\eqref{Delta C in Fourier space}.  This approach allows us to compare our analytical results for the time evolution of the equal-time correlators with those obtained directly from simulations. 

\begin{figure}[!hbt]
\begin{center}  
\includegraphics[trim={0.8cm 0 0cm 0.0cm},clip,width=.99\columnwidth]{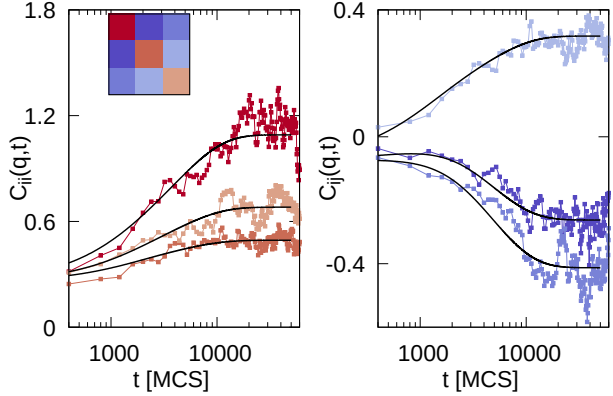}
\caption{Time evolution of the $3\times 3$ correlation matrix in Fourier space after a quench from $T_i=\infty$ to $T=0.9$ at density $\rho=0.7$ and with uniform concentrations $c_i=1/3$, from kinetic Monte Carlo simulations on a square lattice of area $N=150^2$. Top: diagonal elements, bottom: off-diagonal elements (see inset for colour code of entries of $\mathbb{C}$). We used the kinetic rule with $w_{pp}=0$ and the interactions were $\epsilon_{11}=1.4$, $\epsilon_{12}=0.8$, $\epsilon_{13}=0.6$, $\epsilon_{22}=1.0$, $\epsilon_{23}=1.3$ and $\epsilon_{33}=1.2$. Mobilities $\mathbb{L}$ were extracted from a single-component reference fluid with interaction strength $\epsilon=1$ while the effective interaction $\mathbb{\alpha}$ 
was obtained from equilibrium simulations of the mixture at the final temperature $T$. The solid lines are the corresponding analytical predictions for $\mathbb{C}$.
}
    \label{fig:correlation030303}
  \end{center}  
\end{figure}

\begin{figure}[hbt]
  \begin{center}  
    \includegraphics[trim={0.7cm 0 0cm 0.0cm},clip,width=.99\columnwidth]{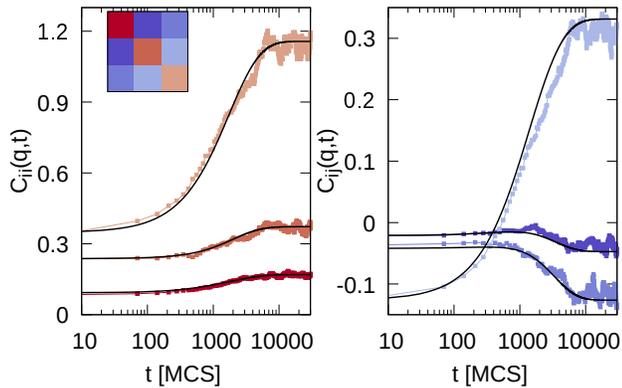}
    \caption{Same as Fig.~\ref{fig:correlation030303} but with $c_1=0.1$, $c_2=0.3$ and $c_3=0.6$.}
    \label{fig:correlation010306}
  \end{center}  
\end{figure}

In Figs.~\ref{fig:correlation030303} and~\ref{fig:correlation010306} we show the time evolution of the equal-time correlation function in Fourier space for different sets of concentrations, averaged over $k=25$ initial conditions. The almost perfect agreement supports our main hypothesis that the transient kinetics of a multi-component mixture can be described with information extracted solely from an equilibrium, single-component reference fluid. In particular, since the same values of $\hat{S}$, $S_0$ and $\hat{S}^s$ were used for different sets of concentrations, we conclude that our general expression for the mobility in Eq.~\eqref{mobilityPainted} correctly expresses the dependency of $\mathbb{L}$ on the species concentrations in the mixture.

\begin{figure}[hbt]
  \begin{center}  
    \includegraphics[trim={1.0cm 0 0cm 0.0cm},clip,width=.99\columnwidth]{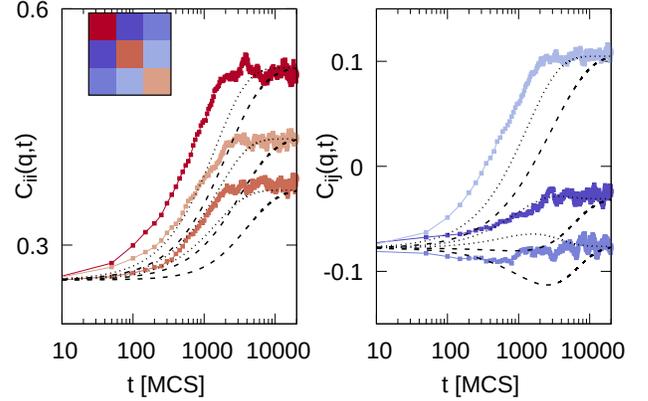}
    \caption{Time evolution of the $3\times 3$ correlation matrix in Fourier space after a quench from $T_i=\infty$ to $T=0.9$ at density $\rho=0.7$ and with uniform concentrations $c_i=1/3$. We used a kinetic rule allowing particle-particle swaps with rate $w_{pp}=0.2w_{pv}$. Interactions were $\epsilon_{11}=1.2$, $\epsilon_{12}=0.9$, $\epsilon_{13}=0.8$, $\epsilon_{22}=1.0$, $\epsilon_{23}=1.15$ and $\epsilon_{33}=1.1$. Dashed (dotted) lines: analytical predictions with standard mobility matrices defined by $r=1$ ($r=1-\rho$). The quantitative and qualitative mismatch between simulations and theory with an incorrect mobility shows the importance of correctly determining the structure of the mobility matrix, especially the value of the parameter $r$. For this plot, to enhance statistics, we averaged over $700$ initial configurations.}
    \label{fig:correlationMixed}
  \end{center}  
\end{figure}

\begin{figure}[hbt]
  \begin{center}  
    \includegraphics[trim={1.0cm 0 0cm 0.0cm},clip,width=.99\columnwidth]{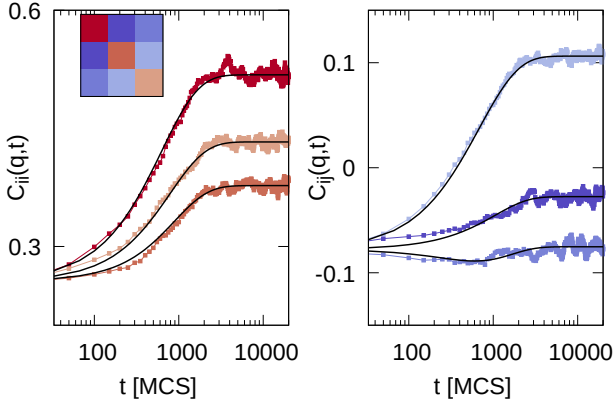}
    \caption{Same as Fig.~\ref{fig:correlationMixed} but with the mobility derived in this work, i.e. Eq.~\eqref{mobilityPainted}. As seen, Eq.~\eqref{mobilityPainted} yields very accurate results, capturing correctly the timescales and prefactors of the correlation matrix.}
    \label{fig:correlationMixed2}
  \end{center}  
\end{figure}

Finally, to illustrate the importance of the underlying kinetics for the mobility and thus the time evolution of the correlation matrix, we also performed simulations with mixed kinetic rules, namely $w_{pp}=0.2w_{pv}$ and compare the measured $\mathbb{C}(q,t)$ with analytical predictions obtained using standard mobilities found in the literature. As Fig.~\ref{fig:correlationMixed} shows, not only are the timescales quantitatively shifted when such mobilities are used, but also the qualitative behavior of the  correlations is no longer captured. While the simulations show a nonmonotonic increasing of only one off-diagonal term, the use of standard mobilities wrongly predicts this behavior for two of them, at the same time it misses the relative strengths of the correlation prefactors for the different timescales. In Fig.~\ref{fig:correlationMixed2} we show the same data is perfectly described once we use the mobilities predicted by our painted particle model. This again stresses the importance of the underlying kinetic rules in determining the mobilities which, in turn, have a noticeable effect on the time evolution of correlations.

\section{Conclusion} 
\label{sec:concl}

In this paper, our aim was to describe the kinetics of multi-component fluids as encoded in the corresponding mobility matrix. Using the hypothesis that kinetic properties depend only mildly on thermodynamic properties such as species-specific interactions, we introduce a model where all particles are identical, being distinguishable only by a color. We then proceed to obtain the mobility for this system and reduce the original matrix problem of finding $\mathcal{O}(M^2)$ quantities to the problem of extracting three scalars from a single-component fluid, which can be easily done via scattering experiments, by performing particle-based simulations or by using theoretical approaches such as Mode Coupling Theory~\cite{Janssen2018}. Even though the input into our mobility comes from a single-component system, it has a non-trivial matrix structure that explicitly reflects the mixture composition. This matrix structure is tuned by a single dimensionless parameter $r$, and for simple choices of $r$ retrieves standard mobilities from the literature. 

By investigating the properties of the mobility we identify two different regimes, dominated respectively by collective motion and interdiffusion, corresponding to fluctuations of the total density or of the local composition. The crossover from one case to the other depends on the microscopic details of the mixture as encoded in the parameter $r$, which folds in on coherent and incoherent structure factor information.

Within a local-in-time approximation, we then applied this mobility to the problem of describing the time evolution of the correlation matrix of a mixture after a quench. We obtain the corresponding time scales arising from the combination of thermodynamics and mobility the interactions, which illustrate the rich behavior of multi-component mixtures. We compare these predictions with the time evolution of the correlation matrix obtained from a Monte Carlo simulation of a hard core lattice gas with nearest neighbor interactions and confirm our kinetic description as well as the validity of our approximation to the mobility. The contrast with predictions from standard mobilities emphasizes the need for careful mobility modelling to capture e.g.\ transient cross-correlations between particle species.

Given the simplicity and broad applicability of our proposed ``painted particle'' mobility matrix, we foresee a broad range of applications in modelling the dynamics of multi-component mixtures. While we have mainly focused on approximating the mobility as local in space and time, it should be emphasized that our approach can in principle retain the full dependence on wavevector $q$ and Laplace rate $z$ of the mobility, and the effects of this will also be fruitful to explore.
On the theoretical side, it will be interesting to analyse in future work the influence of the mobility matrix for quenches into thermodynamically unstable regions. It will also be instructive to investigate the relation of the approach presented here to formally exact methods for the dynamics of Brownian particles \cite{Dean1996,SchmidtBraderJCP_2013_power_func,krugerdean2017b}. 

\section*{Acknowledgments}
  This work was supported by the German Research Foundation (DFG) under grant numbers SO 1790/1-1 and KR 3844/5-1.

\section*{Author Declarations}
\subsection*{Conflict of Interest}
The authors have no conflicts to disclose.

\subsection*{Author Contributions}
\textbf{Maryam Akaberian}: Formal Analysis (equal); investigation (equal); software (equal); visualization (equal); writing - original draft (equal). \textbf{Filipe C Thewes}: Formal Analysis (equal); investigation (equal); software (equal); visualization (equal); writing - original draft (equal). \textbf{Peter Sollich}: Conceptualization (equal); methodology (equal); writing - review and editing (equal). \textbf{Matthias Krüger}: Conceptualization (equal); methodology (equal); writing - review and editing (equal).

\section*{Appendix}
\subsection{Eigenexpansion for real-space correlator}
Here we show how to obtain the real-space correlator Eq.\eqref{correlation in real space} in Sec.~\ref{sec:quench} by writing the correlation matrix from Eq.~\eqref{Delta C in Fourier space} in terms of the eigenvalues and eigenvectors. This enables one to take the Fourier transform and find the expression in real space.
 
We exploit that both  $\Balpha$ and $\mathbb{L}$ are symmetric matrices and expand $\Balpha= \sum_{i}  \alpha _i \boldsymbol {\alpha}_i\boldsymbol{\alpha}_i^\mathsf{T}$ and $\mathbb{\Delta}= \sum_{i}\delta_i \boldsymbol{\delta}_i\boldsymbol {\delta}_i^\mathsf{T}$, where  $\mathbb{\Delta}=T_i\Balpha_i^{-1}-T_f\Balpha^{-1}$. In general, $\mathbb{\Gamma}$ is not a symmetric matrix therefore, we need to distinguish between left and right eigenvectors. We thus use, $\mathbb{\mathbb{\Gamma}}=  \sum_{i} \gamma_i  \boldsymbol{r}_i \boldsymbol{l}_i^\mathsf{T} $, where we assume that $\mathbb{\Gamma}$ can be diagonalized, which holds e.g.\ if the eigenvalues are non-degenerate. We rewrite the correlation matrix in terms of eigenfunction expansion:
\begin{equation*}
    \Delta\mathbb{C}=-\mathbb{\Delta}+\sum_{ik }e^{-q^2(\gamma_i+\gamma_k)t} \boldsymbol{l}_i^\mathsf{T} \mathbb{\Delta} \boldsymbol{l}_k \boldsymbol{r}_i \boldsymbol{r}_k^\mathsf{T}
\end{equation*}
Taking the Fourier transform, we have the correlation matrix in real space
\begin{equation*}\label{correlation in real space app}
    \Delta\mathbb{C}(X,t)=     \sum_{ik}\frac{e^{-\frac{X^2}{4(\gamma_i+\gamma_k)t}}}{[2t(\gamma_i+\gamma_k)]^{\frac{d}{2}}} \boldsymbol {l}_i^\mathsf{T}\mathbb{\Delta}\boldsymbol{l}_k\boldsymbol{r}_i \boldsymbol{r}_k^\mathsf{T}
\end{equation*}

\subsection{Eigenexpansion coefficients}
\begin{figure}[hbt]
  \begin{center}  
    \includegraphics[width=.98\columnwidth]{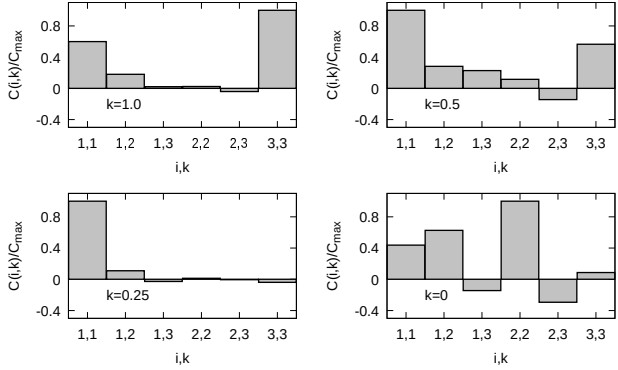}
    \caption{Six independent coefficients of the expansion~\eqref{correlation in real space} for the correlation element $\Delta\mathbb{C}_{11}$. The four different panels correspond to interactions of the form $\epsilon_{ij}=1 + k \chi_{ij}$ where $k=1.0$, $0.5$, $0.25$ and $0$, from top left to bottom right. Here $\chi_{11}=0.4$, $\chi_{12}=-0.2$, $\chi_{13}=-0.4$, $\chi_{22}=0.0$, $\chi_{23}=0.3$ and $\chi_{33}=0.2$.}
    \label{fig:coefficients}
  \end{center}  
\end{figure}

\begin{figure}[hbt]
  \begin{center}  
    \includegraphics[width=.98\columnwidth]{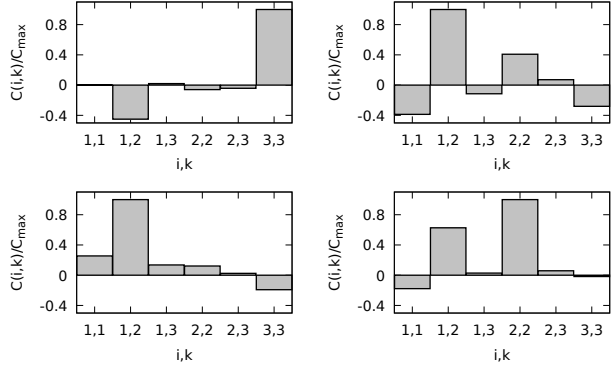}
    \caption{Same as Fig.~\ref{fig:coefficients} but for the off-diagonal element $\Delta C_{13}$. The clear competition between positive and negative coefficients reflects the presence of more than one time scale governing the time evolution of the correlation function.}
    \label{fig:coefficients2}
  \end{center}  
\end{figure}

To show the competition between the different terms in the eigenexpansion above, we compare here the prefactors of the exponentials in~\eqref{correlation in real space}, namely  $[2(\gamma_i+\gamma_k)]^{-\frac{d}{2}} \boldsymbol l_i^\mathsf{T}\mathbb{\Delta}\boldsymbol{l}_k \boldsymbol{r}_i \boldsymbol{r}_k^\mathsf{T}$ for all $i,k$. We choose one diagonal (Fig.~\ref{fig:coefficients}) and one off-diagonal (Fig.~\ref{fig:coefficients2}) element of $\mathbb{C}$. 
 
In the diagonal element, the prefactors are predominantly positive, except for the painted particle interaction, where all timescales become equal and so the individual prefactors become irrelevant (only their sum matters). Therefore, one does not expect a competition between exponentials of different signs of the type we saw in  Sec.~\ref{L-arbitrary}. However, in the off-diagonal element, prefactors of similar magnitude appear, evidencing such a competition. Physically, the kinetics in this case results from an interplay between interdiffusion and collective motion, each at different time scales.

\begin{figure}[hbt]
  \begin{center}  
    \includegraphics[width=.98\columnwidth]{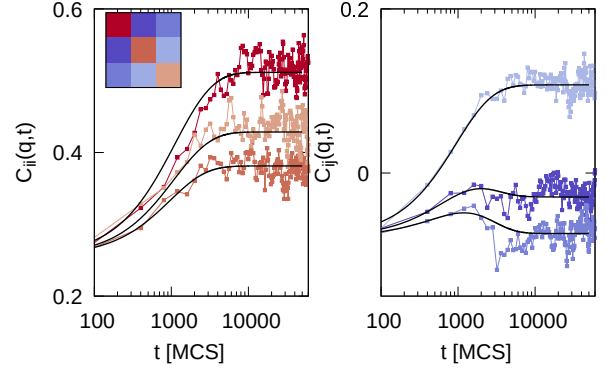}
    \caption{Time evolution of the correlation matrix for interactions as in Fig.~\ref{fig:coefficients} with $k=0.5$. The nonmonotonicity in the off-diagonal elements $C_{12}$ and $C_{13}$ at intermediate times indicates a competition between time scales and physically a transition from collective motion to interdiffusion.}
    \label{fig:evolution2}
  \end{center}  
\end{figure}

In Fig.~\ref{fig:evolution2} we show the time evolution of the correlation function for the case of $k=0.5$ (see caption of Fig.~\ref{fig:coefficients}). Clearly, for the element $C_{13}$ there is a competition between two time scales where, compared to the random initial condition, the correlation first increases and later decreases, indicating an initial collective motion followed by interdiffusion.

\section*{References}
\bibliographystyle{jcp}
\bibliography{references}
\newpage

 \end{document}